\documentclass[3p,times,procedia]{elsarticle}
\usepackage{nupha_ecrc}

\volume{00}

\firstpage{1}

\journalname{Nuclear Physics A}

\runauth{Andrew Connelly, Gregory Johnson, Swagato Mukherjee, and Vladimir Skokov} 

\jid{nupha}

\jnltitlelogo{Nuclear Physics A}

\usepackage{amssymb}

\usepackage[figuresright]{rotating}

\newcommand {\mb} {\mu_B}
\newcommand {\tpc} {T_\mathrm{pc}}
\newcommand {\tc} {T_c^0}
\newcommand {\ms} {m_s^\mathrm{phys}}
\newcommand {\ml} {m_l^\mathrm{phys}}

\newcommand {\zc} {z_c}

\begin{document}

\begin{frontmatter}



\dochead{XXVIIIth International Conference on Ultrarelativistic Nucleus-Nucleus Collisions\\ (Quark Matter 2019)}

\title{
    Universality driven analytic structure of QCD crossover: radius of convergence and QCD critical point
}

%

%
\author[l1]{Andrew Connelly}
\address[l1]{North Carolina State University, Raleigh, NC 27695, USA}
\author[l1]{Gregory Johnson}
\author[l3]{Swagato Mukherjee}
\address[l3]{Brookhaven National Laboratory, Upton, NY 11973, USA}
\author[l1,l2]{Vladimir Skokov}
\address[l2]{Riken-BNL Research Center, Brookhaven National Laboratory, Upton, NY 11973, USA}
\begin{abstract}
Recent lattice QCD calculations show strong indications that the crossover of QCD at zero baryon chemical potential ($\mu_B$) is a remnant of the second order chiral phase transition. The non-universal parameters needed to map temperature $T$ and $\mu_B$ to the universal properties of the second order chiral phase transition were determined by lattice QCD calculations. Motivated by these advances, first, we discuss  the analytic structure of the partition function -- the so-called Yang-Lee edge singularity -- in the QCD crossover regime, solely based on universal properties. Then, utilizing the lattice calculated non-universal parameters, we map this singularity to the real $T$ and complex $\mu_B$ plane, in order to find the radius of convergence for a Taylor series expansion of QCD partition function around $\mu_B=0$ in the QCD crossover regime. Our most important findings are: (i) An universality-based estimate of the radius of convergence  around $\mu_B=0$; (ii) Universality and lattice QCD based constraints on the location of the QCD critical point in the $T-\mu_B$ plane.
\end{abstract}


\end{frontmatter}


\section{Introduction}
\label{Sec:intro}
Lattice QCD calculations  conclusively showed that the approximate chiral symmetry gets nearly restored at a pseudo-critical temperature \(\tpc=156.5 \pm
1.5\)~MeV~\cite{Bazavov:2018mes} via a smooth crossover~\cite{Bhattacharya:2014ara,Bazavov:2011nk}. The same is true also for  small-to-moderate values of \(\mb\) modulo 
the dependence of the transition temperature on the baryon chemical potential \(\tpc(\mb)\)~\cite{Bazavov:2018mes}. 
Beyond this first-principle knowledge, it is {\it conjectured} that at some values of \(\mb\) the chiral restoration in QCD takes place 
via a first order transition (for an alternative, see~Ref.~\cite{Pisarski:2018bct}); the first order phase transition line terminates and turns to crossover at the QCD critical point.

Due to the fermion sign problem, lattice QCD calculations provide limited guidance 
on the existence and location of the QCD critical point
in the \(T-\mb\) phase diagram.
The present lattice calculations can be extended to non-zero \(\mb\) either by carrying out Taylor
expansions around \(\mb=0\)~\cite{Bazavov:2017dus} or through analytic continuation
from purely imaginary values of \(\mb\)~\cite{Bonati:2015bha, Bellwied:2015rza,
Borsanyi:2018grb}. These approaches rely on the assumption that the QCD partition
function is an analytic function of \(\mb\) within a radius of convergence.
To what extent these lattice QCD results are trustworthy, and how far in \(\mb\)
these methods might be extended can be answered only if 
the radius of convergence of the QCD partition around \(\mb=0\) is known.  
In this talk, we use 
non-universal input from lattice QCD calculations to map 
O(4) universality class scaling functions and its singularity structure to   
the QCD \(T\) and \(\mu\) plane and extract the radius of convergence.  


\begin{figure}[!t]
    \centering
    \includegraphics[width=0.5\textwidth]{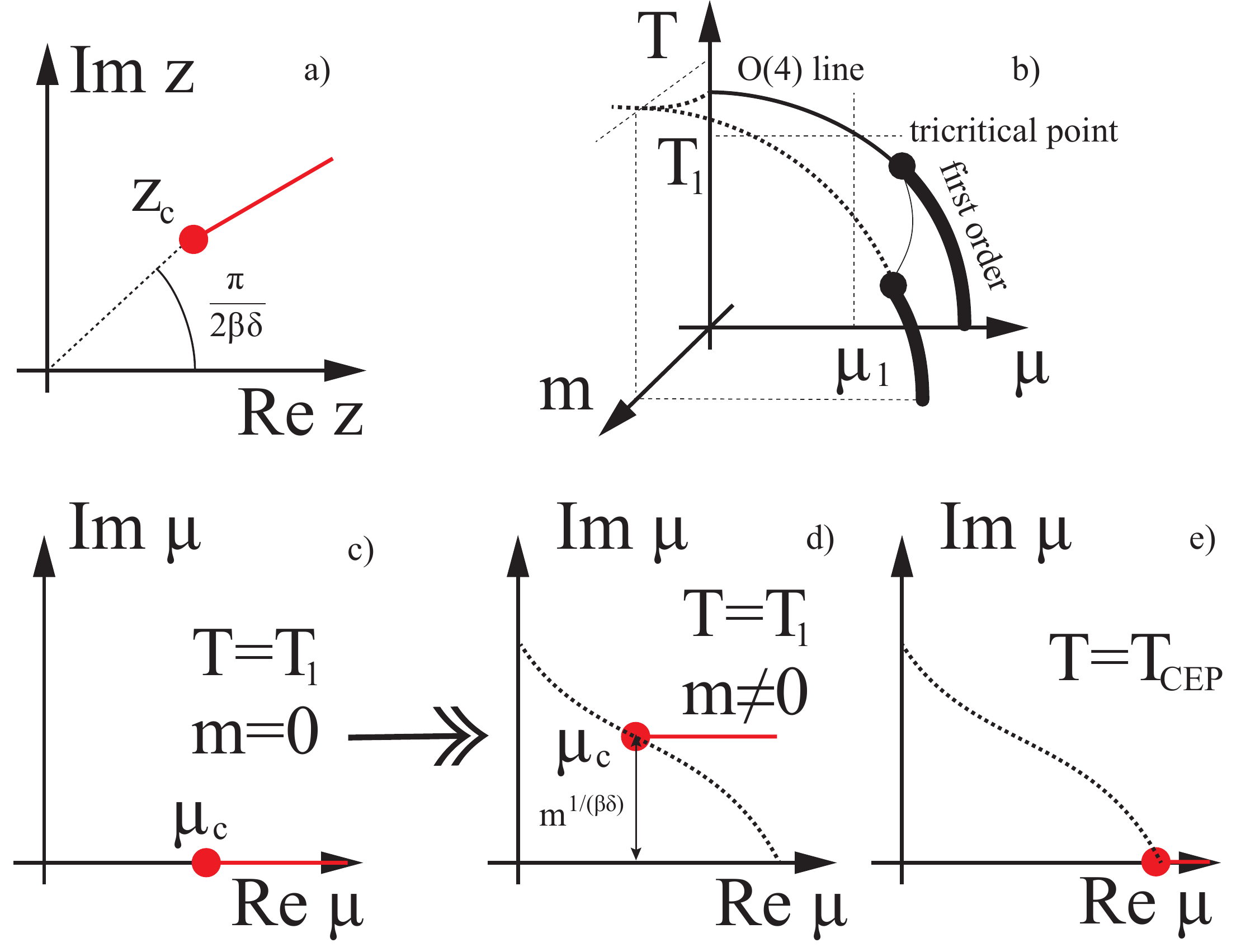}
    \caption{Illustration of singularity structure: a) Yang-Lee edge singularity of the magnetic equation of state; the argument of the singularity is defined by the O(4) critical exponents; b)  illustration of phase diagram in the plane of  temperature -- baryon chemical potential -- light quark mass; c) analytic structure in the chiral limit for $T_1>T_{\rm tricritical}$; d)  analytic structure for non-zero light quark mass for $T_1>T_{\rm CEP}$;  e)  the same as d) but for  $T=T_{\rm CEP}$.  O(4) magnetic equation of state with mapping~(\ref{eq:z}) fairly describes analytical structure in the O(4) scaling regime.}  
    \label{fig:Ill}
\end{figure}

\section{Results and Discussions}
\label{Sec:Res}
Lattice QCD provides a compelling evidence that for massless u and d quarks, the chiral symmetry restoration in QCD takes place via true second order phase transition belonging to the O(4) universality class~\footnote{It is not completely ruled out that the phase transition is of a first order in the chiral limit; it turns into a second order phase transition  of the  three-dimensional Z(2) universality class at some small but non-zero light quark mass. If it is indeed the case, the mapping in Eq.~(\ref{eq:z}) has to be modified, but the main idea still holds.}. The ``external field'' explicitly breaking O(4) symmetry $h$ is related to the mass of the light u and d quarks, conventionally $h=m_l/\ms$.   
Then, for a non-zero $h$, but within the O(4) scaling regime the thermodynamics of QCD is driven by the free energy scaling function $f_s = h^{1+1/\delta} f(z)$ and the magnetic equation of state for the order parameter  $M \equiv \partial f_s / \partial h = h^{1/\delta} g(z)$. Here $z$ is the scaling variable $z =  h^{-1/\beta\delta} t$ and $t$ is the reduced temperature; $z$ can be related to physical $T$ and $\mu_B$ via: 
\begin{equation}
  z = z_0 \left( \frac{m_l}{\ms}\right)^{-\frac{1}{\beta\delta}} 
  \left[ \frac{T-\tc}{\tc} + \kappa^B_2 \left( \frac{\mb}{\tc} \right)^2 +  \kappa^B_4 \left( \frac{\mb}{\tc}  \right)^4  + \dots \right] \,.
  \label{eq:z}
\end{equation} 
In this parametrization, we have a few non-universal parameters: (i) $T_c^0$
is the critical temperature of the chiral phase transition in the limit $m_l=0$; (ii) $\kappa^B_{2,4}$ defines the curvature 
of the transition line at non-zero chemical potential; (iii) $z_0$ is a non-universal constant fixed by universal behaviour of the order parameter as a function of the light quark mass.   These non-universal parameters 
were determined in LQCD calculations: $T_c = 132^{+3}_{-6}$ MeV~\cite{Ding:2019prx,Ding:2018auz}, $\kappa^B_2=0.012(2)$~\cite{Bazavov:2018mes},
$z_0=1-2$~\cite{Bazavov:2018mes}.  LQCD found that $\kappa^B_4$ is consistent with 0.  
The ratio of light to heavy quark masses is  $m_{l}/\ms=1/27$ at the physical point.   
We approximate $z_0$ as a constant. As a non-universal parameter,
$z_0$ can, in principal, have a 
residual dependence on quark mass, temperature and chemical potential. 

It is very well known that for an O($N$) universality class, the functions $g(z)$ and $f(z)$ have a singularity in the complex $z$ plane, see Fig.~\ref{fig:Ill}a. This is the so called Yang-Lee edge singularity --  the remnant of the second order phase transition. The singularity can be treated as an ordinary critical point; it belongs to Z(2) universality class of $\phi^3$ theory.  The symmetry of the partition function with respect to $h\to-h$ and $h\to h^*$ dictates that $z_c = |z_c| \exp \frac{ i \pi }{2 \beta \delta}$ with O(4) critical exponents.   
The magnitude $|z_c|$, the main ingredient in defining the radius of convergence,  was unknown before our studies except for the mean-field approximation or in the large $N$ limit.  
For pedagogical demonstration let's consider the former. The Landau mean-field model for the phase transition 
is 
$    \Omega  = \frac{1}{2} t \sigma^2 + \frac{1}{4}  \sigma^4 - h \sigma \,, 
$
where without any loss of generality we set the quartic coefficient to 1/4 (this fixes $z_0$ to 1). 
The equation of motion for the $\sigma$ field is thus 
\begin{equation}
    \frac{ \partial \Omega}{\partial \sigma}  = t \sigma  +  \sigma^3 - h   = 0.   
\end{equation} 
Introducing $M=h^{1/3} g$ (that is $\delta$ in mean-field approximation is 3) and $z = t /h^{2/3}$ (which also suggests that $\beta=1/2$) this equation  can be rewritten in a canonical form  
\begin{equation}
    g(z) \left [ z +  g^2(z) \right ]   = 1.   
    \label{Eq:g}
\end{equation} 
The position of the singularity is given by the zero of the derivative of the inverse function $z'(g_c) =0$. In combination with Eq.~(\ref{Eq:g}), this gives
$	z_c  = \frac{3} { 2^{2/3} } e^{ \pm i \frac{\pi} {3} } \,.$

For this talk, we performed Functional Renormalization Group (FRG) to establish the value of $z_c$ beyond the mean-field approximation. FRG is based on a functional differential equation for a scale dependent effective action which begins with a bare classical action and iteratively incorporates quantum fluctuations by the momentum type scale resulting in the fully renormalized action. A standard basic approach to solving the FRG flow equations for the effective action is to take the lowest order derivative expansion of the effective action together with a truncated Taylor series for the effective potential. One can extract $g(z)$ from the effective potential and plot its imaginary part as a function of $|z|$ and Arg$(z)$. FRG fails in the vicinity of the Yang-Lee edge singularity, and we estimated $|z_c| $ as the nearest point before failure giving $|z_c|\approx 1.68$.

From the equation $z = z_c$, one can determine the position of the Yang-Lee edge singularity 
in the complex chemical potential plane, $\mu_c$, at a given $T$. The absolute 
value of $\mu_c$ defines the radius
of convergence of the Taylor series expansion of pressure in powers of $\mu$, 
that is $R_\mu = |\mu_c| $. This assumes
that the Yang-Lee edge singularity is the closest to the origin. 

\begin{figure}[!t]
    \centering
    \includegraphics[width=0.49\textwidth]{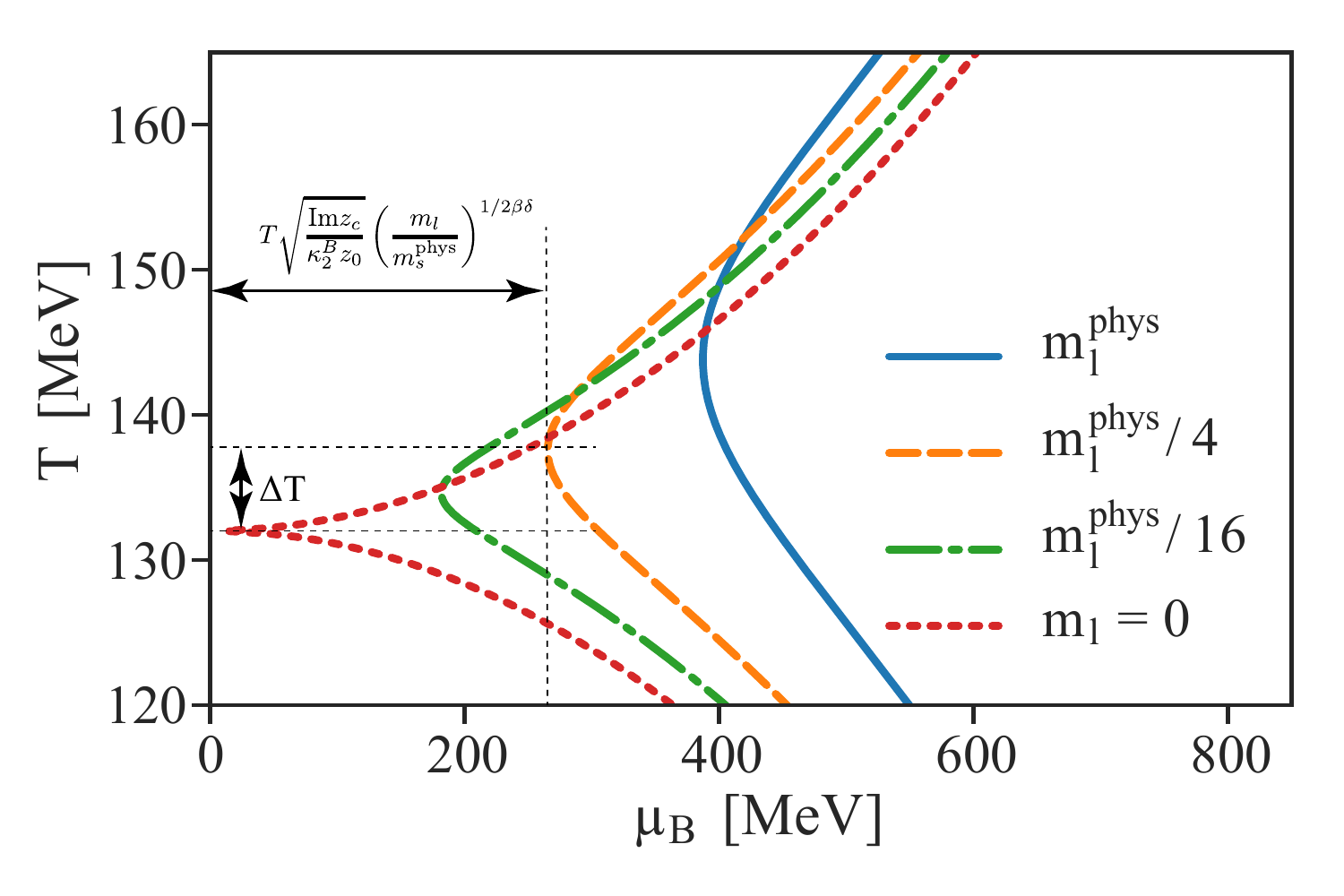}
    \includegraphics[width=0.49\textwidth]{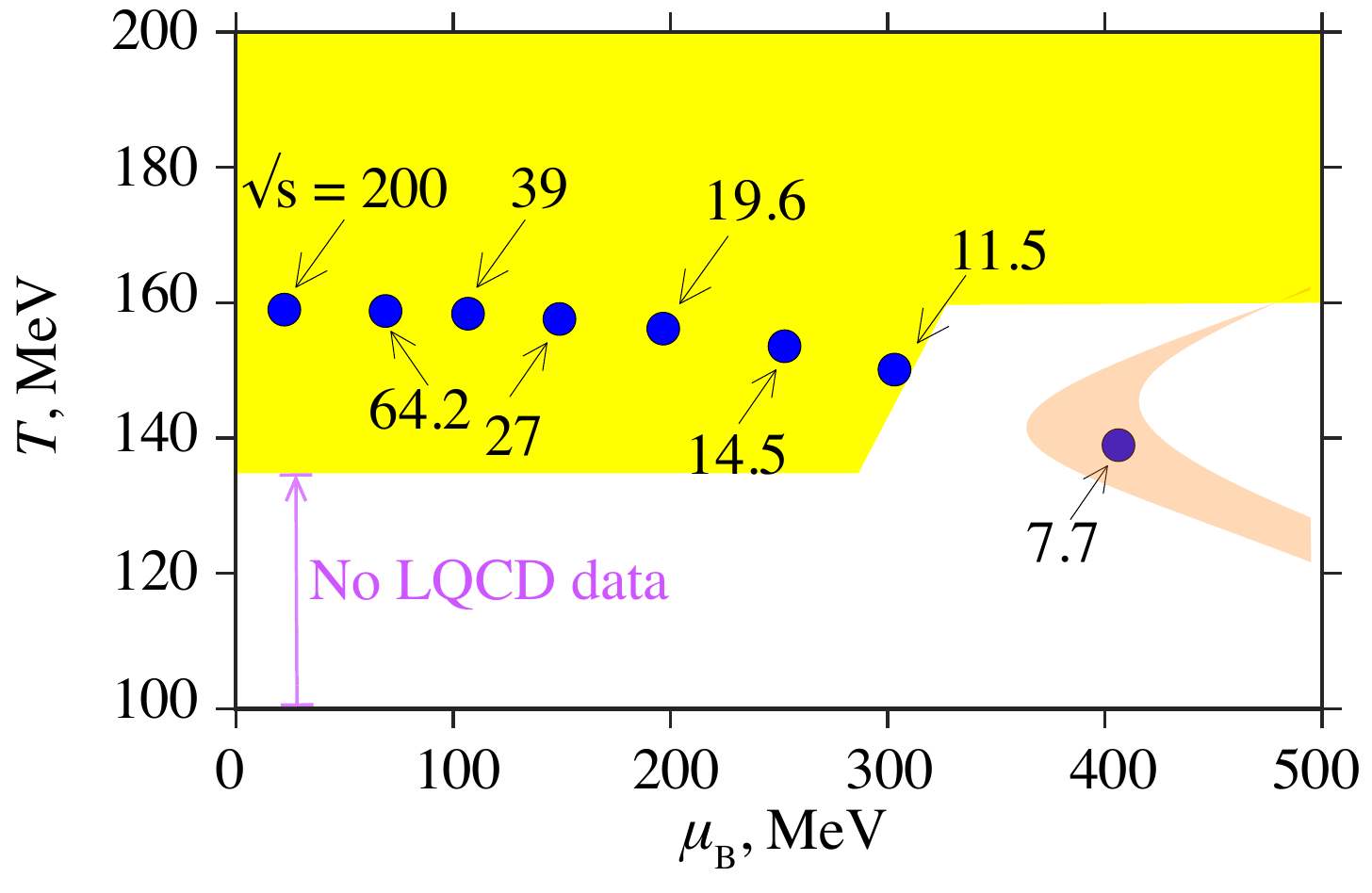}
    \caption{{\bf Left panel}: Radius of convergence \(\mb\) as a function for different values of the
    light up/down quark masses. The minimum of the curves shifts to higher temperatures  
    by the amount $\frac{\Delta T}{T^0_c} = \frac{{\rm Re} z_c} {z_0}  \left( \frac{ m_l }{ m_s} \right)^{\frac{1}{\beta\delta}}$.  {\bf Right panel}: Radius of convergence in \(\mb\) for physical quark masses. The orange
    band is for {\(z_0=2\)} and incorporates a 15\% uncertainty on the value of \(|\zc|\). 
    The yellow region 
    depicts LQCD disfavoured region for the lacation of crtitical end point. The blue dots show freeze-out $T$ and $\mu_B$ for a given collisions energy in GeV.   
    }  
    \label{fig:Runc}
\end{figure}

In Figure~\ref{fig:Runc}, left panel,  we show the radius of convergence in \(\mb\) in the \(T-\mb\)
plane for different values of \(m_l=0-\ml\), using  \(z_0=2\), \(O(4)\) critical exponents, 
and other lattice QCD-determined
non-universal parameters described above. Note that, in the chiral limit, QCD free
energy is singular at \(T=\tc\), \(\mb=0\) and, therefore, the radius of
convergence at this point is zero. 

Figure~\ref{fig:Runc}, right panel,  provides a more realistic estimate for the radius of convergence
in \(\mb\) in the \(T-\mb\) plane for \(\ml\) by varying \(|\zc|\) around its FRG value; as to account the  systematics related to the FRG truncation used in our calculations.   
While the variation of
\(|\zc|\) leads to a limited uncertainty of the radius of convergence, more precise
lattice QCD results for \(z_0\) are needed to improve this estimate.

\section{Conclusions}
\label{Sec:Conc}

Relying only on the universal behavior of QCD in the chiral crossover region we
investigated the analytic behavior of the free energy. We argued that if the chiral
behavior of QCD is well-described by the universal scaling, as borne out in recent
the lattice QCD calculations, then analytical structure of the free energy will be completely
governed by  the corresponding universal scaling function in
the complex scaling variable. We extracted the position of  the relevant singularity of the scaling
function by performing FRG  calculations. 
We showed
how this can be translated to the singularity in the complex-\(\mb\) plane
to determine the radius of convergence in \(\mb\), solely based on the universal
critical exponents and well-determined non-universal parameters from lattice QCD
calculations. Figure~\ref{fig:Runc} summarizes our universality- and QCD-based
estimate for the radius of convergence in \(\mb\) for temperatures in the vicinity of
the QCD chiral crossover. This shows that the radius of convergence is larger than
\(|\mb| \gtrsim 400\)~MeV, implying that the present lattice QCD calculations based on
Taylor expansions in \(\mb\) and analytic continuations from imaginary values of
\(\mb\) can be reliable below this region, as suggested also by recent lattice QCD
calculations~\cite{Bazavov:2018mes, Bazavov:2017dus, Borsanyi:2019hsj}.

The present state-of-the-art lattice QCD calculations do not find any evidence for an
additional singularity for \(\mb \lesssim 400\)~MeV~\cite{Bazavov:2018mes,
Bazavov:2017dus,  Borsanyi:2019hsj}. Our result on the radius of convergence
\(|\mb|\gtrsim400\)~MeV,  coupled with these lattice QCD results, suggest that QCD
critical point, if one exists, will most likely be located at \(\mb\gtrsim400\)~MeV.
Such conclusion will potentially have an important impact on the on-going beam energy
scan experiments at RHIC and SPS, as well as on the future experiments, such as at
FAIR and NICA.  {Since the
critical point is located somewhere along the chiral crossover boundary,
naturally, the corresponding singularity is continuously connected to the crossover Yang-Lee edge 
singularity. In fact, a critical point is located where  the Yang-Lee edge singularity and
its complex conjugate pinch the real chemical potential
axis~\cite{Stephanov:2006dn}.} Thus, the curves in Figure~\ref{fig:Runc} help
understand how to map critical Ising direction, \(t\), to QCD, which is of relevance
not only for static~\cite{Pradeep:2019ccv} but also dynamic
properties~\cite{Martinez:2019bsn} near a possible critical point.

This work was supported by the U.S. Department of Energy, Office of Science, Office of Nuclear Physics: (2) Through the Contracts No. DE-SC0012704 and DE-SC0020081; (iii) Within the framework of the Beam Energy Scan Theory (BEST) Topical Collaboration. 

\bibliographystyle{elsarticle-num}
\bibliography{ref}

\end{document}